\documentclass[12pt,preprint]{emulateapj}

\usepackage[usenames,graphicx]{}
\usepackage{color}

\slugcomment{}

\shorttitle{Time-dependent Corotation Resonance in Barred Galaxies}
\shortauthors{Wu et al.}

\begin{document}

\title{Time-dependent Corotation Resonance in Barred Galaxies}

\author{Yu-Ting Wu\altaffilmark{1}, Daniel Pfenniger\altaffilmark{2}  and
  Ronald E. Taam\altaffilmark{1,3}}
\affil{$^1$Institute of Astronomy and Astrophysics, Academia Sinica, Taipei 10617, Taiwan\\
$^2$Geneva Observatory, University of Geneva, CH-1290 Sauverny, Switzerland\\
$^3$Department of Physics and Astronomy, Northwestern University, 2145 Sheridan Road, Evanston, IL 60208, USA}  

\begin{abstract}
  The effective potential neighboring the corotation resonance region in barred galaxies is shown to be strongly
  time-dependent in any rotating frame because of the competition of nearby perturbations of similar strengths with differing
    rotation speeds.  Contrary to the generally adopted assumption, that in the bar rotating frame the corotation region should
  possess four stationary equilibrium points (Lagrange points), with high quality $N$-body simulations we localize the instantaneous
  equilibrium points and find that they circulate or oscillate broadly in azimuth with respect to the pattern speeds of the inner or
  outer perturbations.  This implies that at the particle level the Jacobi integral is not well conserved around the corotation
  radius. That is, angular momentum exchanges decouple from energy exchanges, enhancing the chaotic diffusion of stars through the
  corotation region.
\end{abstract}


\section{Introduction}

Most spiral galaxies, including the Milky Way, are barred with the interface between the central bar and its surrounding spiral
region corresponding to the corotation region, a broad circular annulus showing a progressive transition between the bar and the
spiral arms.  This region marks the transition between two fundamentally different dynamical regimes: in the bar region energy and
angular momentum are exchanged at the maximally possible rate given by the bar rotation rate \citep{PfeFri91}, while in the disk
region these quantities become progressively better conserved at larger radii. Angular momentum is increasingly better conserved
when the potential is closer to rotational symmetry, while energy is increasingly better conserved as the potential becomes
time-invariant in an inertial frame.

Dynamical studies in disk galaxies have, since the 1960's, been focused on the spiral density wave theory pioneered by
\cite{LinShu64}, and \cite{GolLyn65} where the spirals are considered as small perturbations of an axisymmetric background
potential, and are supposed to rotate with a constant pattern speed.  In this particular rotating frame, circular orbit
resonances exist, among which the corotation resonance is the most important one: it marks the region of the galaxy where the star
orbital period is close to the spiral perturbation period.  The corotation resonance is characterized by a number of stationary
Lagrange points in the rotating frame which characterize the dynamics around the corotation region.  Of these points, the unstable
ones generate chaos which is the ultimate cause of the irreversible behavior linked with resonances.

Starting in the 1970's \citep{deVFre72,Con73} dynamical studies of bars adopted the same conceptual framework as in the spiral
density wave theory.  Specifically, the bar has been considered as a perturbation of the background potential with a well defined
corotation resonance. This assumption is implicit in most publications on the topic \citep[e.g.,][Sect.\ 3.3.2]{BinTre08}, but
remained unchecked to date.  However, bars are much stronger non-linear perturbations than spirals, and their dynamical influence
ranges well beyond the corotation region, typically inducing spiral arms which rotate at a different, slower pattern speed than the
bar itself \citep{SelSpa88}.  In this context, it is therefore not obvious that the conceptual framework adopted for a weak spiral
perturbation in the linear approximation remains valid since the bar and outer spiral perturbation are both strong and rotate at
different frequencies.  It is the purpose of this paper to show that the location of the equilibrium points in self-consistent
models are essentially time-dependent, and that the commonly assumed dynamics in the corotation region is not well described by a
theoretical model based on stationary Lagrange points.

\section{Models}

To examine the dynamics of the corotation region of a barred galaxy, we use 3D $N$-body models based on the currently most accurate
method to set up initial conditions for equilibrium multi-component galaxy models \citep[GalIC:][]{YouSpr14} and perform high
quality, momentum conserving pure stellar dynamical model simulations \citep[gyrfalcON:][]{Deh00}.  The details are not essential
for the discussion here, but will be described elsewhere \citep{WuPfeTaa16}.

The initial galaxy model is constructed using three axisymmetric Miyamoto-Nagai (hereafter MN) components \citep{MiyNag75}
corresponding to a bulge, a disk, and a halo with a chosen $Q$-parameter profile.  The parameters of these three
  components are listed in Table~\ref{table:3MN}.  

The model develops two long lived nested bars surrounded by an outer spiral region.  The basic rule we have followed for producing
long-lived double bar systems is to start the simulations with initial equilibrium axisymmetric models having a double-peaked
rotation curve with equal maxima, using two of the three MN components, fixing the respective bar sizes, with the smaller component
having a colder kinematics ($Q \sim 1$) than the medium component ($Q\sim 1.5$), as in \cite{DuEtal15}. The bulk of each bar shows a
well defined pattern speed, while the spiral region is described by a distinct non-uniform pattern speed, decreasing with increasing
radius.  The third MN component is spherical (a Plummer model) with a large core, and is used mainly to keep the rotation curve high
in the outer spiral region.

With $N= 2\times10^7$, the particle noise is sufficiently low to unambiguously find the instantaneous rotation speed of each
bar, and the instantaneous locations of the equilibrium points (hereafter EPs), defined by the locations where the acceleration vanishes
in the respective rotating frames. Each of these patterns interact with the other adjacent patterns via gravitational
torques, and the regions where these torques compete most effectively are the respective corotation rings of the inner and the outer
bars.

\section{Corotation Dynamics}

\subsection{Evolution of Two Bar Models}
Several figures encompassing several bar--spiral relative phases of a rotating double-barred system allow to illustrate
simultaneously the inner small and outer large bar corotation evolutions.  The motivation for choosing a double-barred system is to
check whether the findings of this paper are different at the interface between two nested bars or at the interface between the
outer bar and its surrounding spiral arms.

In selected snapshots Fig.~\ref{fig:model_015_xy_evo} shows the projected density of the rotating double-barred system including about
0.3\,Gyr from the whole 8\,Gyr simulation, of the inner and outer bars.  The color map indicates the projected surface density in log-scale.
As can be seen from the left panels, where the bars are displayed at 4\,kpc radius scale, the outer and inner bars are aligned at
$t = 3.107\,$Gyr.  They rotate counter-clockwise.  Because the inner bar rotates faster than the outer bar, the angle between them
changes with time. At $t = 3.116\,$Gyr the angle is about $45^{\circ}$, and at $t = 3.126\,$Gyr the angle is $90^{\circ}$.
Alignment occurs again at $t = 3.146\,$Gyr.

The structure revealed in the right panels at 20\,kpc radius scale clearly shows the outer bar and its outer spiral arm pattern,
mainly shown in yellow and red; they are aligned at $t = 2.975\,$Gyr. Due to their different rotation speeds, they become
perpendicular to each other at $t =3.179\,$Gyr and are again aligned with each other at $t = 3.295\,$Gyr.

\subsection{Pattern Speeds}
To investigate quantitatively the dynamics in the vicinity of the corotation region of a given bar, we need the bar pattern rotation
frequency $\Omega_\mathrm{b}$ around the rotation axis $z$.  We expect that each bar and spiral may be mutually perturbed, so the
pattern speed of each structure may oscillate in time.

There are several methods to determine a pattern speed from a set of particles \citep{PfeSahWu16}.  We retain here a straightforward
method based on the polar moment of inertia tensor. For a set of particles of mass $m_i$ and Cartesian coordinates $x_i$, $y_i$
whose center of mass is at the origin the polar moment of inertia tensor in the $(x-y)$-plane reads
\begin{equation}
  \mathbf{I}
  \equiv 
  \left(
    \begin{array}{cc}
     I_{xx}  & I_{xy}\\
      I_{xy} &  I_{yy}
   \end{array}
  \right)
  \equiv 
  \left(
    \begin{array}{cc}
     \sum_i m_i x_i^2   & \sum_i m_i x_i y_i\\
      \sum_i m_i x_i y_i & \sum_i m_i  y_i^2
   \end{array}
  \right). 
\end{equation}
The selected particles of a given bar belong to a ring included in the bulk of the bar region.  The precise size of the ring is not
important as long as it includes the bulk of the measured bar.  Here, the inner bar ring is taken between 0.1 and 0.4\,kpc in
  radius, and the outer bar ring is between 2.5 and 4.5\,kpc. This allows one to calculate the azimuthal angle $\phi_\mathrm{b}$
of the bar major axis from the eigenvectors direction of the tensor $\mathbf{I}$, yielding, after some algebra,
\footnote{The single argument arctan formula given by \cite{BinMer98} [Sect.\ 10.3.2] in the mathematically similar context of the
  vertex deviation delivers angles over only a quadrant due to the factor $1/2$, preventing to track continuously the same principal
  axis. }
\begin{equation}
  \phi_\mathrm{b}(t_k)=
  \frac{1}{2}\mathop{\mathrm{arctan2}} \left(2I_{xy},I_{xx}-I_{yy}\right).
\end{equation}
The pattern speed of the bar can be measured by estimating the time-derivative of this angle, which is here determined by an average
centered finite difference over two time-intervals of $\pm 2\,$Myr before and after the evaluated time to filter out the highest
frequencies.

\subsection{Instantaneous Equilibrium Points}

Given the structures associated with the inner bar, outer bar and outer spiral, we now focus on the properties of the
instantaneous EPs.  The EPs in the frame rotating at the known frequency $\Omega_\mathrm{b}$ are found
by finding the extrema of the effective potential
\begin{equation}
  \Phi_\mathrm{eff} = \Phi - {1\over 2} \Omega_\mathrm{b}^2 R^2, 
\end{equation}
where $R^2= x^2+y^2$, and $\Phi$ is the gravitational potential.  
This can be achieved by finding the zero acceleration points
including the centrifugal force component,
\begin{eqnarray}
  0 &=& a_R + \Omega_\mathrm{b}^2 R = {a_x x + a_y y \over R} +
  \Omega_\mathrm{b}^2 R , \label{eq-cir_freq_scalar}\\
  0 &=& a_\phi = {a_x y - a_y x \over R} , \label{eq-a_phi_0}\\
  0 &=& a_z.
\end{eqnarray}
where $\vec a = -\nabla \Phi$, and $a_{k}$ are the respective gravitational acceleration components in Cartesian or polar
coordinates. To find the respective EPs in the inner bar and the outer bar rotating frame, the pattern speed of each bar is
  used for $\Omega_\mathrm{b}$.

\begin{center}
\begin{figure}[t!]
  \epsscale{0.7}
 \plotone{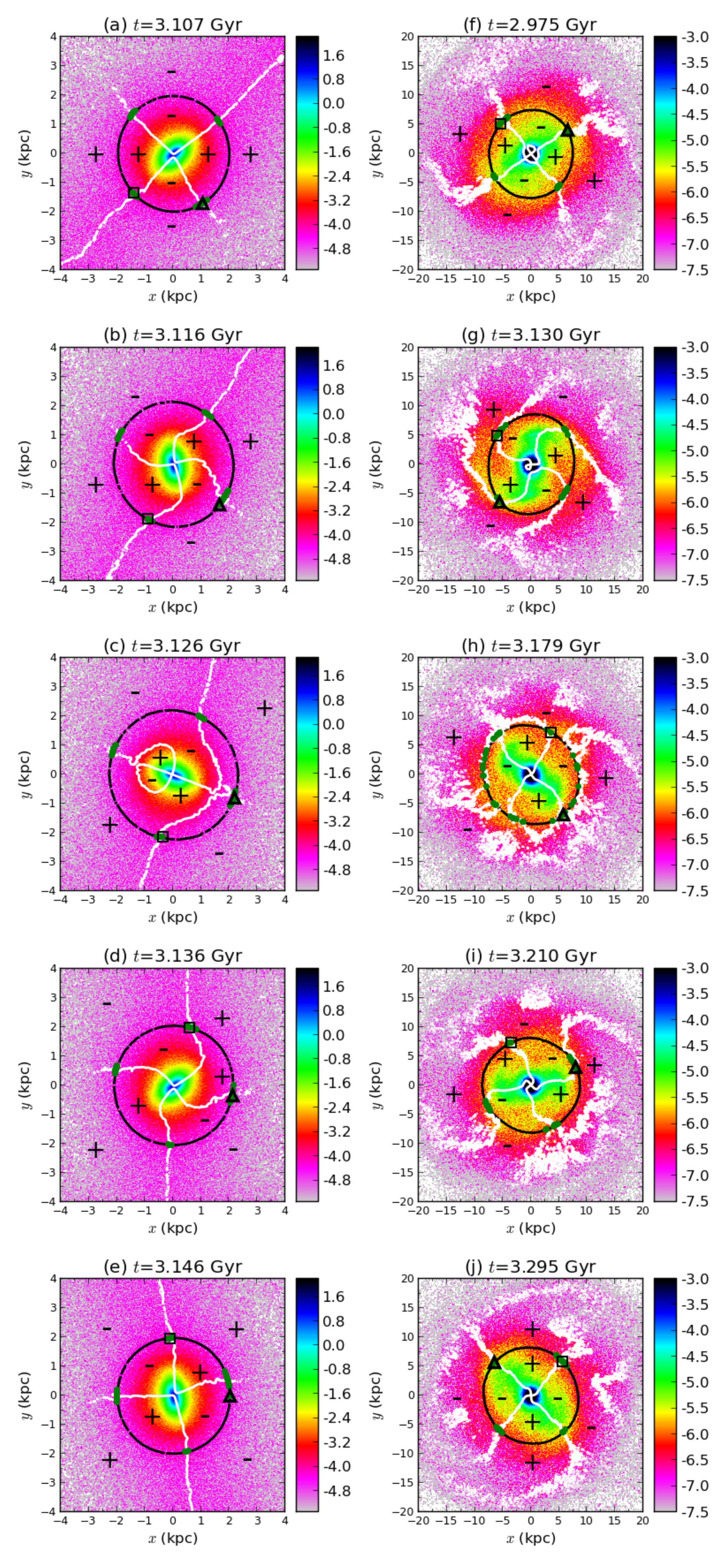}
 \caption{Evolution of the inner bar (left panel (a)-(e)) and the outer bar projected density (right panel (f)-(j)) at selected
   times.  The black and white curves show the zero radial and tangential accelerations, respectively (satisfying
   Eq.~(\ref{eq-cir_freq_scalar}-\ref{eq-a_phi_0})).  The green points show the equilibrium point locations.  The black square
     and black triangle show the location of the equilibrium points with the smallest and the largest radius, respectively.  The
   $+$ and $-$ indicate the torque sign.}
\label {fig:model_015_xy_evo}
\end{figure}
\end{center}

The EPs can be visually inspected by the intersections of the black curves and the white curves in
Fig.~\ref{fig:model_015_xy_evo}.  These curves show the locations satisfying the zero-acceleration conditions
Eqs.~(\ref{eq-cir_freq_scalar}-\ref{eq-a_phi_0}), respectively.  Specifically, the black curves in Fig.~\ref{fig:model_015_xy_evo}
represent the corotation ovals where the radial acceleration vanishes, and the white curves show the zero-torque locations.

To determine the EPs, we adopt the following procedure.  For each bar the particles located within $|z|<$ half of the disk
  scale height are binned in a $x-y$ grid (pixel size of $\Delta x = 0.03$ and 0.09\,kpc for the inner and outer bar grids) and the
summed squares of the right-hand sides of Eqs.~(\ref{eq-cir_freq_scalar}-\ref{eq-a_phi_0}) times $R$
\footnote{The $a_z=0$ condition holds anyway within $|z|< \Delta z$, where $\Delta z$ is about half the disk
  scale-height, because the disks are never significantly warped.  The multiplication by $R$ of the horizontal acceleration is used
  to avoid numerical overflow near the origin.  },
\begin{equation}
  W = \left({a_x x + a_y y} + \Omega_\mathrm{b}^2 R^2\right)^2 + \left( a_x y - a_y x\right)^2
   \label{eq:W}
\end{equation}
is averaged in the grid cells. 
The disk scale height is 0.45 kpc here, as listed in Table~\ref{table:3MN}.
 For each snapshot, the 36 gridded pixels which are not close to the origin and contain the smallest
values of $W$ are selected as a sample of points spatially close to the true EPs.  These pixels are represented by the green points
in Fig.~\ref{fig:model_015_xy_evo}.  At any given time, a different number of pixels may be found near any EPs, but over time the
evolution of the cloud of points provides an approximate estimate of the EP positions.  The number 36 is retained based on the fact
that in average 9 points for 4 EPs sample well the nearest pixels of the sought points.

From the left panels of Fig.~\ref{fig:model_015_xy_evo}, it is evident that the EPs near the inner bar are mainly
affected by the outer bar and lie close to the semi-major and semi-minor axis of the outer bar.  However, because the outer bar
rotates with respect to the inner bar, the EPs are not stationary in the inner bar rotating frame.

On a larger spatial scale, corresponding to the right panels of Fig.~\ref{fig:model_015_xy_evo}, the EPs between the outer bar and
the spiral arms are shown.  It is apparent that the spiral structure causes a torque which competes with the bar torque, resulting
in the bending of the white lines, especially just outside the corotation region represented by the black curves.  When the outer
bar and its outer spiral structure are in phase, as shown in panel (f), the white curves are bent beyond $R>10\,$kpc, where the
torque is dominated by the outer spiral structure.  Here, the EPs are located along the semi-major and semi-minor axis of the outer
bar. Subsequently, since the rotation frequencies of the outer bar and its surrounding spiral arms differ, the two structures become
out of phase as shown in panel (g).  During this stage, the white curves are bent near the corotation region and the EPs, as
indicated by the green points, are no longer near the semi-major and semi-minor axis of the outer bar.  When the two structures are
out of phase by about $90^{\circ}$, as shown in panel (h), the white curves are bent and overlay the corotation region.  Therefore,
the EPs are located at arbitrary angles near the corotation region.  We note that, at this stage, the direction of the torque, as
indicated by the $+$ and $-$ signs, differs between inside and outside the corotation region.  For example, along $y=0$ axis, the
torque is negative inside the corotation region, but positive outside the corotation region.  Later, the two structures are in phase
again, as shown in panel (j), and the EPs are located near the semi-major and semi-minor axis of the outer bar, similar in
appearance to panel (f), with the directions of the torque unchanged inside and outside the corotation region.

In summary, given that the contributions to the torque around corotation from the inner and outer non-axisymmetric structures
are similar, the instantaneous EPs move in any rotating frame. Thus, except for the potential minimum at the origin, there are no
stationary points in any rotating frame. In other words, no true stationary Lagrange points exist.

\subsection{Time Evolution of Equilibrium Points}
In this subsection, we describe quantitatively the temporal variation of the equilibrium point parameters. For the corotation
region of the inner and outer bar, their radii as well as their azimuths oscillate substantially.  Specifically, Fig.~\ref{fig:RTime} and
Fig.~\ref{fig:phiTime} show the radius and the reduced azimuthal angle (described in section 3.4.2) of the EPs as functions of
time.  In both figures, the dashed and dotted vertical lines mark the times when two bars are aligned and perpendicular to each
other, respectively.

\subsubsection{Equilibrium Point Radii}
\begin{center}
\begin{figure}[t!]
  \vspace{-3cm}
  \epsscale{1.3}
  \plotone{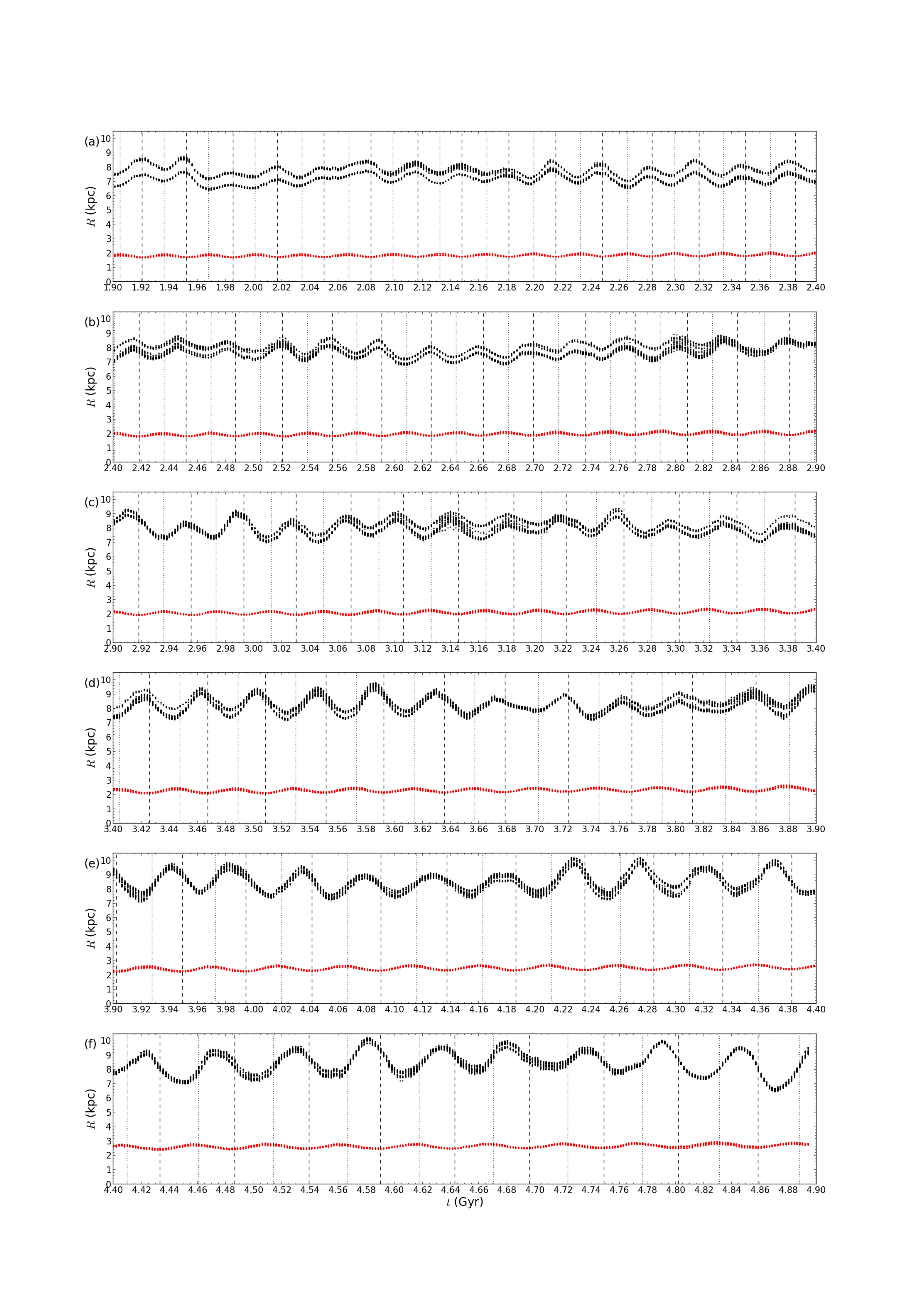}
  \vspace{-2.5cm}
  \caption{Time evolution of the equilibrium point radii over 3\,Gyr after the formation of the double-barred system.  At any time,
    the positions of the true equilibrium points are sampled by 36 dots with the lowest $W$ values (Eq.~(\ref{eq:W})).  The red and
    black dots are for the inner and outer corotation region, respectively.  The dashed and dotted lines mark the times when two
    bars are aligned and perpendicular to each other, respectively.}
\label {fig:RTime}
\end{figure}
\end{center}

\begin{center}
\begin{figure}[!h]
  \vspace{-3cm}
  \epsscale{1.3}
  \plotone{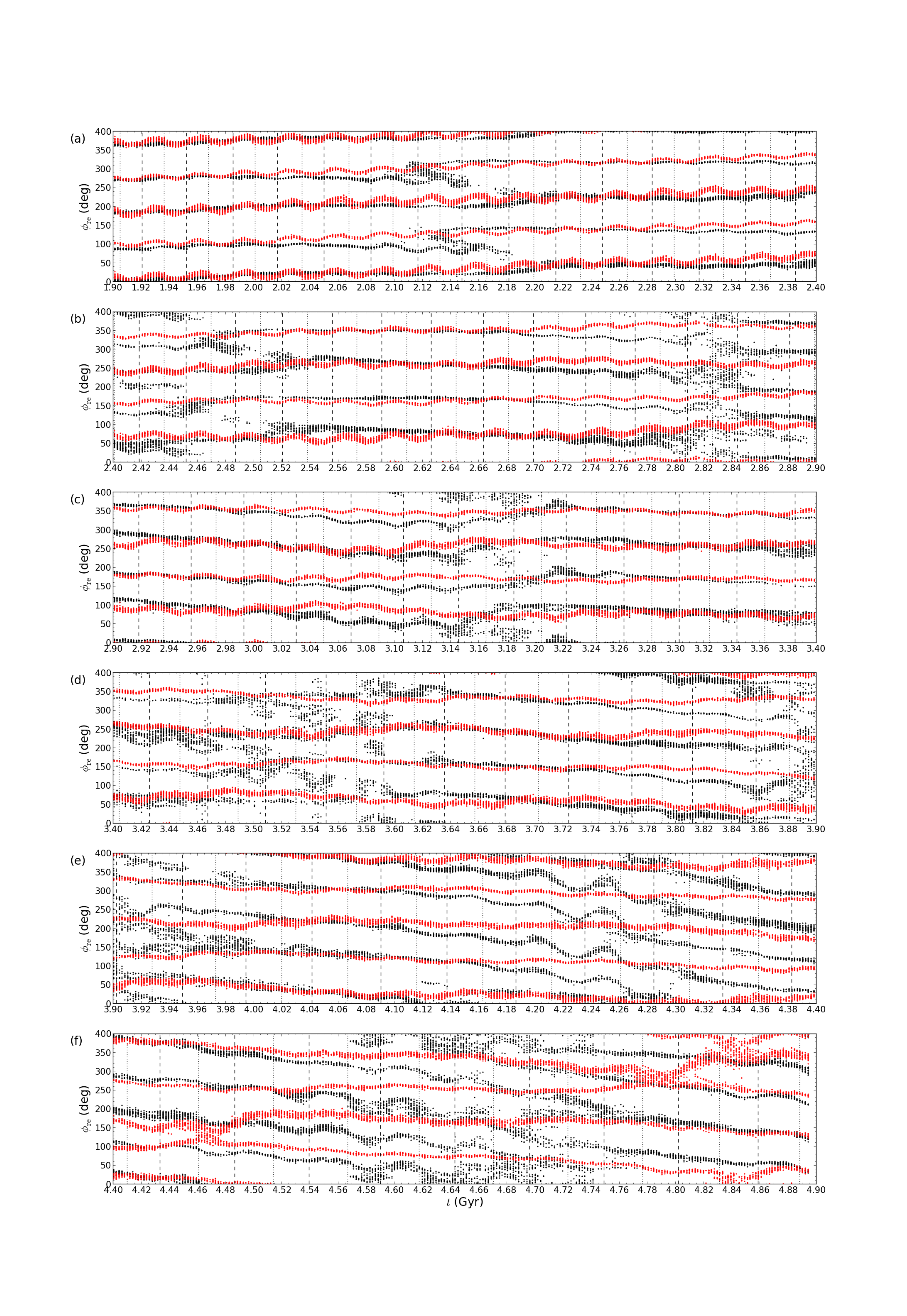}
  \vspace{-2.5cm}
  \caption{Time evolution of the equilibrium point azimuths. The dots and lines have the same meaning as in
    Fig.~\ref{fig:RTime}. The dots more than $360^\circ$ are repeated dots between $0$ - $360^\circ$.}
\label {fig:phiTime}
\end{figure}
\end{center}

From Fig.~\ref{fig:RTime}, it is clear that the radii of the EPs of the inner corotation region (hereafter ICR) and the outer
corotation region (hereafter OCR) oscillate substantially ($\sim \pm 5-10\%$) according to the azimuth difference between the inner
and outer bar. Most of time, the EPs can be classified into two groups according to their radii.  Namely, for the EPs of the OCR,
the group with a larger radius is located near the bar major axis, whereas the group characterized by a smaller radius is located
near the bar minor axis.  In Fig.~\ref{fig:model_015_xy_evo} (f)--(j), the black square and black triangle mark the location of the
EPs with the smallest and the largest radius, respectively.  However, for the EPs of the ICR, the group with larger radius are
located along the semi-minor axis of the outer bar, and the group with smaller radius are located along the semi-major axis of the
outer bar.  Again, the black square and black triangle in Fig.~\ref{fig:model_015_xy_evo} (a)--(e) show the location of the EPs with
the smallest and the largest radius, respectively.

The maximum and the minimum radii of the EPs of the ICR are in phase with the relative angle of the two bars indicated by the dotted
and dashed lines in Fig.~\ref{fig:RTime}, while the maximum and the minimum radii of the EPs of the OCR are sometimes shifted by
more than $30^\circ$. This is due to the interaction of the spiral structure external to the outer bar.

At different times, the EPs of the OCR no longer exhibit a bimodal distribution in the radial range, such as
at $t=2.80\,\rm Gyr$.  This corresponds to events involving the phase difference between the outer bar and spiral structure and
will be described in more detail below.

\subsubsection{Equilibrium Point Azimuths}
To facilitate the presentation of the time-evolution of the EPs azimuths $\phi_i$, which are monotonously growing
functions of time, in Fig.~\ref{fig:phiTime} we plot the reduced azimuth $\phi_\mathrm{re}$,
\begin{equation}
 \phi_\mathrm{re} = \phi_i - \bar\Omega (t-t_o),
\label{eq:phi_re}
\end{equation}
where $\bar\Omega = 25 \,\rm km \, s^{-1} \, kpc^{-1}$ is a constant frequency, which is similar to the outer bar frequency over the analyzed 
time interval, and $t_o$ is the time of the starting point.  The reduced azimuth plot allows
one to perceive more easily the relative
changes of angular velocity.  When the EPs rotate faster/slower, the curve slope increases/decreases.  For example,
the EPs of the ICR (red dots) rotate faster than $\bar\Omega$ at $t = 2.05\,\rm Gyr$ and rotate slower at $t
= 2.07\,\rm Gyr$.

The EPs azimuths of the ICR (red points) are split into four distinct groups, for example at $\phi_\mathrm{re}= 10$, 100, 190,
280$^\circ$ at $t=1.90\,\rm Gyr$, which is similar to Lagrange points except that they oscillate. Their regular periodic oscillation
is in phase with the relative angle of the two bars: they rotate faster/slower when the bars are aligned/perpendicular.  The typical
oscillation amplitude amounts to $\pm 10^\circ$ within a period equal to $\pi/(\Omega_\mathrm{i} -\Omega_\mathrm{o})$, where
$\Omega_\mathrm{i}$ and $\Omega_\mathrm{o}$ are the inner and outer bar pattern speeds, respectively.  Departures from such a
description are seen in the last frame (f), during short episodes where two of these groups shortly merge showing that further
complications can occur.

The EP azimuths of the OCR (black points) are also split into four distinct groups with oscillation amplitudes of
$\sim \pm 10^\circ$ within a period $\pi/(\Omega_\mathrm{i} -\Omega_\mathrm{o})$. However, about every $0.3-0.4\,\rm Gyr$, they are distributed over all angles ($\sim \pm 90^\circ$) due to $\sim 90^\circ$ phase difference between the outer bar and the spiral
structure, such as at $t=2.83$, 3.18$\,\rm Gyr$.  Fig.\ 1 (h) shows a typical snapshot of this behavior.  In addition, when the bars
are aligned or perpendicular to each other, the EPs of the OCR rotate at approximate constant speed since the slope of
$\phi_\mathrm{re}$ there is almost constant.

Overall, the average slope of the EPs of the ICR is slightly larger than the EPs of the OCR, showing that the EPs of the ICR rotate
with the speed of the outer bar, as well as the EPs of the OCR, except that occasionally the EPs of the OCR sharply decrease their
speed during the episodes when they are distributed over all angles as described in the previous paragraph.  During these episodes,
the corotation resonance loses its strength as discussed in the next subsection.

\subsubsection{The Effective Potential }

\begin{center}
\begin{figure}[!h]
  \vspace{-3cm}
  \epsscale{1.3}
  \plotone{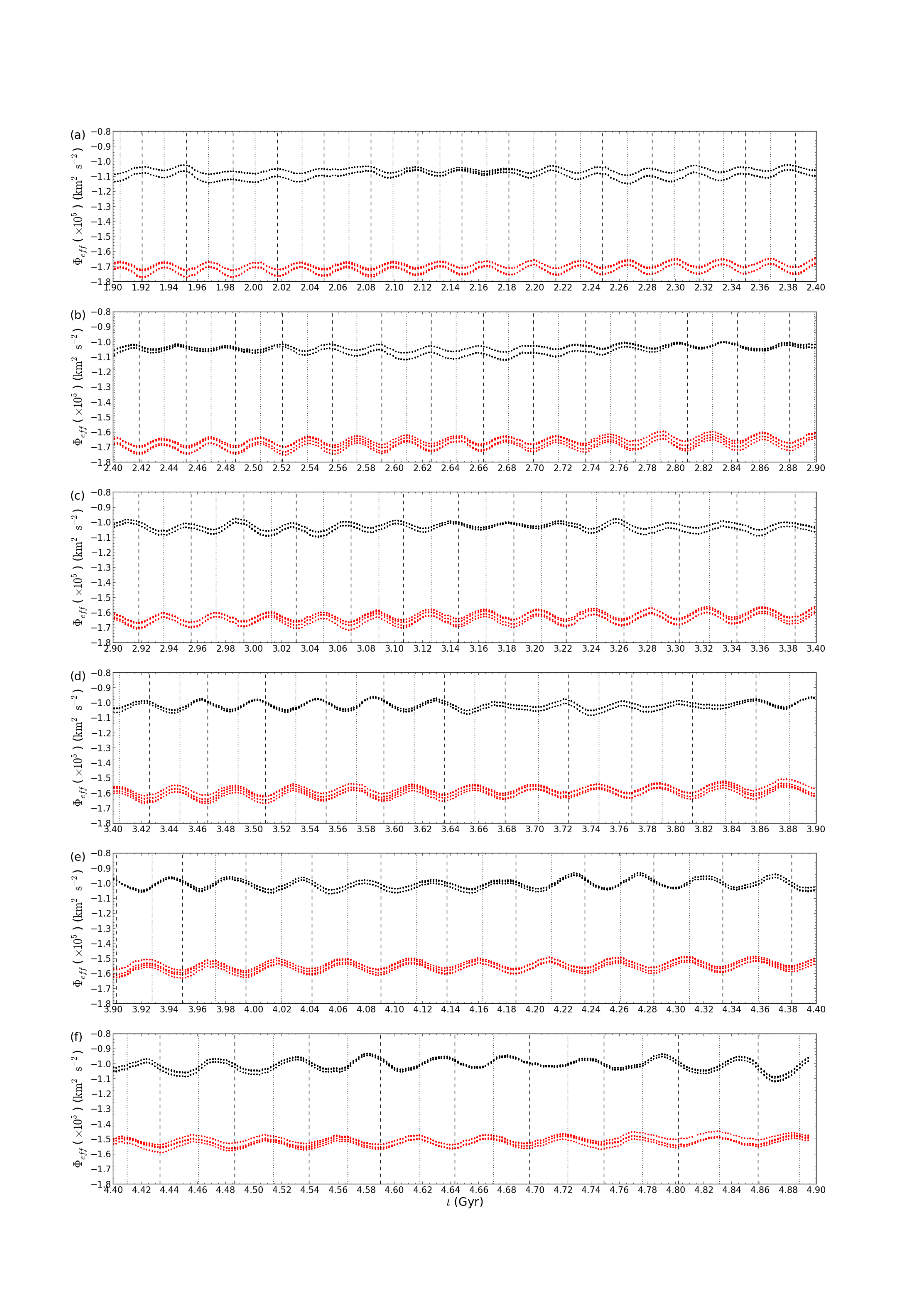}
  \vspace{-2.5cm}
  \caption{Time evolution of the effective potential $\Phi_\mathrm{eff}$ at the equilibrium points. The dots and lines have the same
    meaning as in Fig.~\ref{fig:RTime}.}
  \label {fig:effPotTime}
\end{figure}
\end{center}

As a consequence of the temporal radial and azimuthal variations described above, there is a corresponding time-dependence of the
effective potential $\Phi_\mathrm{eff} = \Phi - {1\over 2} \Omega_\mathrm{b}^2 R^2$ at the EPs.  Specifically,
Fig.~\ref{fig:effPotTime} shows the time-dependence of the effective potential at the EPs.  As in Fig.~\ref{fig:phiTime}, the dashed
and dotted lines represent the times when the bars are aligned and perpendicular to each other, respectively. The red dots denote
the EPs of the ICR, while the black dots denote the EPs of the OCR.  For both kinds of dots, the lower values correspond to the
local $\Phi_\mathrm{eff}$ saddle points, and the higher values correspond to the local $\Phi_\mathrm{eff}$ maxima.

For the EPs of the ICR and OCR, the amplitude of the oscillations amounts to $\sim\pm 5\%$ over almost all the shown time-interval,
with a period $\pi/(\Omega_\mathrm{i} -\Omega_\mathrm{o})$ as defined in section 3.4.2.  The $\Phi_\mathrm{eff}$ of the EPs of ICR
are in phase with the relative phase of two bars, while over time the $\Phi_\mathrm{eff}$ phase of the EPs of the OCR may lag by up
to 40$^\circ$, which results from the bar-spiral interactions.

For the EPs of the OCR, the variations between the saddle points and the maxima of $\Phi_\mathrm{eff}$ at any time $t$, i.e.
$\Delta \Phi_\mathrm{eff}(t) = \Phi_\mathrm{eff}^\mathrm{max}(t) -\Phi_\mathrm{eff}^\mathrm{saddle}(t)$, oscillate between $0$ to
$\sim \pm 2\%$. At some times, $\Delta \Phi_\mathrm{eff}$ becomes very small, such as at $t=2.83$ or 3.18$\,\rm Gyr$. Over
time-scale of order $\sim 100\,\rm Myr$, the OCR becomes less effective. This feature repeats with a period of $300-400\, \rm
Myr$. Such variations modulate the strength of the corotation resonance, and its influence on the outer disk dynamics.

Comparing the $\Delta \Phi_\mathrm{eff}(t)$ at any given time ($0$ to $\sim \pm 2\%$) and the amplitude variation over a period
$\pi/(\Omega_\mathrm{i} -\Omega_\mathrm{o})$ ($\sim\pm 5\%$), the time variations of $\Phi_\mathrm{eff}$ over a period, which is
comparable with the bar rotation times, are more important than $\Delta \Phi_\mathrm{eff} (t)$ over the corotation region at any
 specific time. This demonstrates that, unlike true Lagrange points, these EPs are essentially time-dependent.

What is not perceptible in Figs.\ $2-4$ is that for the EPs of the ICR, the saddle points of $\Phi_\mathrm{eff}$ are located at
slightly smaller radii than the maxima of $\Phi_\mathrm{eff}$ (by about $0.1\,$kpc), which is the opposite from the EPs radii of the
OCR (or the classical Lagrange points in bars), where the saddle points are located at larger radii than the maxima of
$\Phi_\mathrm{eff}$ (by about $0.6\,$kpc).

\section{Conclusions}

The results of this paper is a logical corollary of the \cite{SelSpa88} work.  In their work the pattern speeds of the bar and
spiral arms have been studied and found to differ, which implies that the respective patterns torque each other, and cannot be
strictly rigid.  Thus, around the corotation regions, time-independence is impossible in any rotating frame.

We have analyzed the dynamical consequences of both the bar-bar and bar-spiral pattern interactions quantitatively in $N$-body
simulations of a typical galaxy disk model, including two nested bars and spiral arms.  We found that the common assumptions that 1)
rigid bar, and 2) the existence of stationary equilibrium points in a particular rotating frame, are invalidated by the high degree
of time-dependence in these systems.

The dynamics in the two corotation regions is shown to be essentially time-dependent because the instantaneous equilibrium points
oscillate with a period comparable to the orbital period, over distances ($1-2\,\rm kpc$) larger than the width of the corotation
regions. Further, over similar periods, the effective potential values at these equilibrium points are modulated by a larger amount
than the differences between their maxima and minima at any given time. This demonstrates the strong time-dependence of the
  potential and the non-existence of true Lagrange points. 

In usual bar models, where the bar is supposed to be rigid and rotating with a constant speed $\Omega_\mathrm{b}$, the Hamiltonian
$H = E - \Omega_\mathrm{b} L_z$ is time-independent ($E=v^2/2 + \Phi$ is the particle energy and $L_z$ the particle angular
momentum), so $H$ is an integral of motion, the so-called Jacobi integral.  In our simulation, since $\Phi$ is time-dependent in any
rotating frame, $H$ is time-dependent. While $E$ and $L_z$ still vary, they are no longer constrained to conserve $H$.  Therefore,
the consequence is that chaotic motion must be much more widespread than in the time-independent case, and the effect of such a kind
of corotation resonance is much more diffusive than estimated in analytic theories which assumes time-independent dynamics.  This
provides a dynamical understanding of the results in \cite{BruChi11}, where in a barred galaxy model the radial migration of stars,
measured by a radial diffusion rate, was found to be the largest around the bar corotation radius, and not in the spiral region. We
conclude that the \cite{SelBin02} migration scenario attributed to transient spirals must be supplemented by an eventually larger
contribution to stellar migration when a bar is present, since the bar-spiral corotation region is strongly time-dependent.

We have seen that the equilibrium points oscillate in azimuth back and forth up to $40^\circ$, which means that the tips of the bars
are twisted to similar extent (see Fig.\ 1(g)). In other words, bars are actually rather flexible structures, especially near their
ends.  Near their ends, the assumptions of rigidity and constant pattern speed may be improper. At any particular time, a bar may
present a differently twisted pattern, and therefore we expect that estimations of the Milky Way bar pattern speed made with star
samples located near its corotation radius (at $4-6\,\rm kpc$ from the Galactic Center) might give a different pattern speed than
estimations derived from star samples located well within the bulk of the bar. In the future, to assure the Milky Way basic
parameters, such as the bar pattern speed or the bar orientation, we will need to take into account the particular twisting stage of
the Milky Way bar.

\acknowledgments

The authors are grateful to W.~Dehnen for the gyfalcON code and D.~Yurin \& V.~Springel for the GalIC code.  The authors also thank
the support of the Theoretical Institute for Advanced Research in Astrophysics (TIARA) based in Academia Sinica Institute of
Astronomy and Astrophysics (ASIAA) and Sam Tseng for assistance on the TIARA computational facilities and resources.  The authors
would like to acknowledge the support of the Geneva Observatory and its computational facilities, as well as Yves Revaz for his
assistance.


\begin{table}[h]   
\begin{center}
  \caption[Parameters of three Miyamoto-Nagai components.]
          {Parameters of three MN components \citep[notations as in][]{MiyNag75}}
 \begin{tabular}{lcccc}
 \hline
 \hline
 \multicolumn{2}{l}{Parameter}& Halo & Disk & Bulge \\
 \hline
 mass $M$ ($10^{10} \, \rm M_{\odot}$) && 15.0 & 8.6504 & 1.3496 \\
 scale length $a+b$ (kpc)     && 15.0 & 4.50 & 0.50\\
 scale height $b$ (kpc)       && 15.0 & 0.45 & 0.15\\
 \hline
 \end{tabular}
\label{table:3MN}
\end{center}
\end{table}

\clearpage

\clearpage

\begin{thebibliography}{}

\bibitem[Binney \& Tremaine(2008)]{BinTre08} Binney, J. \& Tremaine,
  S. 2008, ``{Galactic Dynamics: Second Edition}'', Princeton University
  Press, Princeton
  
\bibitem[Binney \& Merrifield(1998)]{BinMer98} Binney, J. \& Merrifield, M.
 1998, ``{Galactic Astronomy}'', Princeton University
  Press, Princeton
  
\bibitem[Brunetti et al.(2011)]{BruChi11}
 Brunetti, M., Chiappini, C., \& Pfenniger, D. 2011, \aap, 534, A75
  
\bibitem[Contopoulos (1973)]{Con73}
  Contopoulos, G. 1973, \apj, 181 657
  
\bibitem[de Vaucouleurs \& Freeman(1972)]{deVFre72}
  de Vaucouleurs G., \& Freeman, K.~C. 1972, Vistas in Astronomy 14, 163

\bibitem[Dehnen(2000)]{Deh00}
  Dehnen, W. 2000, \apjl, 536, L39

\bibitem[Du et al.(2015))]{DuEtal15}
  Du, M., Shen, J., \& Debattista, V. 2015, \apj, 804, 139D
  

\bibitem[Goldreich \& Lynden-Bell(1965)]{GolLyn65}
  Goldreich, P., \& Lynden-Bell, D. 1965, \mnras, 130, 125

\bibitem[Lin \& Shu(1964)]{LinShu64}  
 Lin C.~C., \& Shu, F.~H. 1964, \apj, 140, 646

\bibitem[Miyamoto \& Nagai(1975)]{MiyNag75}
  Miyamoto, M., \& Nagai, R. 1975, \pasj, 27, 533

\bibitem[Pfenniger \& Friedli(1991)]{PfeFri91}
  Pfenniger, D., \& Friedli, D. 1991, \aap, 252, 75

\bibitem[Pfenniger, Kanak \& Wu(2016)]{PfeSahWu16}
  Pfenniger, D., Saha, K., \& Wu, Yu-Ting 2016, in preparation 

\bibitem[Sellwood \& Binney(2002)]{SelBin02}
  Sellwood, J.~A., \& Binney, J.~J. 2002 \mnras, 336, 785
  
\bibitem[Sellwood \& Sparke(1988)]{SelSpa88}
  Sellwood, J.~A., \& Sparke, L.~S. 1988, \mnras, 231, 25P

\bibitem[Yurin \& Springel(2014)]{YouSpr14}
  Yurin, D., \& Springel, V. 2014, \mnras, 444, 62
  
\bibitem[Wu, Pfenniger \& Taam(2016)]{WuPfeTaa16}
  Wu, Yu-Ting, Pfenniger, D., \& Taam, R. 2016, in preparation 

\end{thebibliography}
\end{document}